
\tolerance=10000
\input phyzzx


\REF\nfor{A. Das, M. Fischler and M. Rocek, Phys. Rev. {\bf D16} (1977)
3427.}
\REF\dwn{ B. de Wit and H. Nicolai,  {\it Phys. Lett.} {\bf 108 B} (1982) 285; B.
de Wit and H. Nicolai,  {\it Nucl. Phys.} {\bf B208} (1982) 323.}
\REF\gunwar{ M. G\"unaydin, L.J. Romans and N.P. Warner, {\it Gauged $N=8$
Supergravity in Five Dimensions}, Phys. Lett. {\bf 154B}, n. 4 (1985) 268; {\it
Compact and Non--Compact Gauged Supergravity Theories in Five Dimensions}, Nucl.
Phys. {\bf B272} (1986) 598.}
\REF\PPV{ M. Pernici, K. Pilch and P. van Nieuwenhuizen, {\it Gauged
$N=8~D=5$ Supergravity} Nucl. Phys. {\bf B259} (1985) 460.}
\REF\gat{S.J. Gates and B. Zwiebach,
Nucl.\ Phys.\ B {\bf 238}, 99 (1984);
Phys.\ Lett.\ B {\bf 123}, 200 (1983)
.}
\REF\nct{C.M. Hull, {\it Phys. Rev.} {\bf D30} (1984) 760; C.M. Hull,
{\it Phys. Lett.} {\bf 142B} (1984) 39; C.M. Hull, {\it Phys. Lett.} {\bf
148B} (1984) 297; C.M. Hull, {\it Physica} {\bf 15D} (1985) 230; Nucl.
Phys. {\bf B253} (1985) 650; C.~M.~Hull and N.~P.~Warner,
Nucl.\ Phys.\ B {\bf 253}, 650 (1985)
and 
Nucl.\ Phys.\ B {\bf 253}, 675 (1985).}
\REF\ncto{ C.M. Hull, {\it Physica} {\bf 15D} (1985) 230; Nucl.
Phys. {\bf B253} (1985) 650.}
\REF\nctt{ 
  C.M. Hull, {\it Class. Quant. Grav.} {\bf
2} (1985) 343.}
\REF\CW{C.M. Hull and N. P. Warner, Class. Quant. Grav. {\bf 5} (1988) 1517.}
\REF\DS{C.M. Hull, hep-th/0109213.}
\REF\evacs{
D.~Z.~Freedman and G.~W.~Gibbons,
Nucl.\ Phys.\ B {\bf 233}, 24 (1984);
D.~Z.~Freedman and B.~Zwiebach,
Nucl.\ Phys.\ B {\bf 237}, 573 (1984);
B.~Zwiebach,
Phys.\ Lett.\ B {\bf 135}, 393 (1984).
P.~M.~Cowdall,
Class.\ Quant.\ Grav.\  {\bf 15}, 2937 (1998)
[hep-th/9710214].
}
\REF\randsum{L. Randall and R. Sundrum, Phys. Rev. Lett. {\bf 83} (1999) 3370;
4690.}
\REF\Warn{N.P. Warner, Nucl.\ Phys.\ B {\bf 128}, 169 (1983).}
\REF\Sken{
 K.~Skenderis and P.~K.~Townsend,
hep-th/9909070.}
\REF\cfun{ L. Girardello, M. Petrini, M. Porrati and A. Zaffaroni,
{\sl Novel Local CFT and Exact Results on Perturbations of N=4 Super Yang
Mills from AdS Dynamics}, JHEP {\bf 9812} (1998) 022, hep-th/9810126.}
\REF\Freed{
D.Z. Freedman, S.S. Gubser, K. Pilch and N.P. Warner, {\sl
Renormalization group flows from holography--supersymmetry and a
c-theorem}, hep-th/9904017.}
\REF\LPS{ H. L\"{u}, C.N. Pope, E. Sezgin and K.S. Stelle,
{\sl Dilatonic p-brane solutions}, Phys. Lett. {\bf B371} (1996) 46-50,
hep-th/9511203.}
\REF\BoonstraMP{
H.~J.~Boonstra, K.~Skenderis and P.~K.~Townsend,
JHEP {\bf 9901}, 003 (1999)
[hep-th/9807137].}
\REF\stains{ H. L\"{u}, C.N. Pope, E. Sezgin and K.S. Stelle,
Nucl.  Phys. Lett. {\bf B456} (1995) 669.}
\REF\CowdallTW{
P.~M.~Cowdall, H.~Lu, C.~N.~Pope, K.~S.~Stelle and P.~K.~Townsend,
Nucl.\ Phys.\ B {\bf 486}, 49 (1997)
[hep-th/9608173].}
\REF\LuRH{
H.~Lu, C.~N.~Pope and P.~K.~Townsend,
Phys.\ Lett.\ B {\bf 391}, 39 (1997)
[hep-th/9607164].}
\REF\SinghQD{
H.~Singh,
Phys.\ Lett.\ B {\bf 444}, 327 (1998),
hep-th/9808181.}
\REF\Cveticaa{
M.~Cvetic, S.~Griffies and S.~Rey,
``Static domain walls in N=1 supergravity,''
Nucl.\ Phys.\ B {\bf 381}, 301 (1992)
hep-th/9201007.}
\REF\Cveticbb{
M.~Cvetic and H.~H.~Soleng,
Phys.\ Rept.\  {\bf 282}, 159 (1997), hep-th/9604090.}
\REF\CveticXX{
M.~Cvetic, S.~S.~Gubser, H.~Lu and C.~N.~Pope,
Phys.\ Rev.\ D {\bf 62}, 086003 (2000)
[hep-th/9909121].}
\REF\Behrn{
K.~Behrndt, E.~Bergshoeff, R.~Halbersma and J.~P.~van der Schaar,
Class.\ Quant.\ Grav.\  {\bf 16}, 3517 (1999)
[hep-th/9907006].}
\REF\Ahn{
C.~Ahn and K.~Woo,
hep-th/0109010.}
\REF\Zcoup{
B.~Zwiebach,
Nucl.\ Phys.\ B {\bf 238}, 367 (1984).
}
\REF\FreedS{
D.Z. Freedman and J.H.
 Schwarz, Nucl. Phys. {\bf B137} (1978) 333.}
\REF\CJ{E. Cremmer and B. Julia, Phys. Lett. {\bf 80B} (1978) 48; Nucl.
Phys. {\bf B159} (1979) 141.}
\REF\frefo{ F. Cordaro, P. Fr\'e, L. Gualtieri, P. Termonia and M. Trigiante,
{\it $N=8$ gaugings revisited: an exhaustive classification}, Nucl. Phys. {\bf
B532} (1998) 245.}
\REF\Cremmer{E. Cremmer, in {\it Supergravity and Superspace},
S.W.
Hawking and
M. Ro\v cek, C.U.P.
Cambridge,  1981.}
\REF\fref{
L.~Andrianopoli, F.~Cordaro, P.~Fre and L.~Gualtieri,
Fortsch.\ Phys.\  {\bf 49}, 511 (2001)
[arXiv:hep-th/0012203]; C.M.Hull, unpublished.
}
\REF\PPVa{ M. Pernici, K. Pilch and P. van Nieuwenhuizen, Nucl. Phys. {\bf
B249} (1985) 381.}


\font\mybb=msbm10 at 12pt
\def\bbbb#1{\hbox{\mybb#1}}

\def\R {\bbbb{R}}
\def\bE{\bbbb{E}}
 %

\def \aa {\alpha}

\def \dd {\delta}
\def \ee {\epsilon}

\def \ll {\lambda}

 \def \ggg {\Gamma}

\def \lll {\Lambda}

\def \2 {{1 \over 2}}
\def \3 {{1 \over 3}}
\def \4 {{1 \over 4}}
\def \5 {{1 \over 5}}
\def \6 {{1 \over 6}}
\def \7 {{1 \over 7}}
\def \8 {{1 \over 8}}
\def \9 {{1 \over 9}}
\def \0 { \infty}

\def\++ {{(+)}}
\def \- {{(-)}}
\def\+-{{(\pm)}}

\def\ek {\eqn\abc$$}

\def \pa {\partial}

\def \qq {\qquad}
\def\vvv{{\varphi}}

 \def\unit{\hbox to 3.3pt{\hskip1.3pt \vrule height 7pt width .4pt \hskip.7pt
\vrule height 7.85pt width .4pt \kern-2.4pt
\hrulefill \kern-3pt
\raise 4pt\hbox{\char'40}}}

\def\nup#1({Nucl.\ Phys.\  {\bf B#1}\ (}



\Pubnum{ \vbox{  \hbox {QMUL-PH-01-11}  
\hbox{hep-th/0110048}
}}
\pubtype{}
\date{October 2001}

\titlepage

\title {\bf  Domain Wall and 
de Sitter Solutions of Gauged Supergravity}

\author{C.M. Hull}
\address{Physics Department,
\break
Queen Mary, University of London,
\break
Mile End Road, London E1 4NS, U.K.}
\vskip 0.5cm

\abstract {BPS domain wall solutions of gauged supergravities
are found, including those theories which have non-compact gauge groups.
 These include models that have both an unstable de Sitter solution and
stable domain wall solutions.
}

\endpage

\chapter{Introduction}

Scalar potentials whose dependence on one of the scalars fields $\phi$
is of the form
$$V=\lll -  \aa (1-\cosh (a  \phi   ))
\eqn\lampot$$
occur in many supergravity theories, with extrema of $V$ giving
anti-de Sitter or de Sitter solutions with cosmological constant $\lll$.
There are many examples with $\lll <0$, such as the 
$D=4$, $N=4$ gauged supergravity of [\nfor], the
$D=4$, $N=8$ gauged supergravity with gauge group $SO(8)$
[\dwn] and the
$D=5$, $N=8$ gauged supergravity with gauge group $SO(6)$ [\gunwar,\PPV].
The case $\lll>0$ also occurs, for example in the
$D=4$, $N=4$ gauged supergravity of [\gat], the
$D=4$, $N=8$ gauged supergravity with gauge group $SO(4,4)$
[\nct-\nctt] and the
$D=5$, $N=8$ gauged supergravity with gauge group $SO(3,3)$ [\gunwar].
These solutions were lifted to solutions of supergravity in 10 or 11
dimensions in [\CW]; see [\DS] for further discussion. The supergravities
with 
$\lll<0$ typically    
have stable maximally  supersymmetric AdS vacua, but the
de Sitter solutions arising when $\lll>0$ necessarily break all
supersymmetries, and moreover are unstable as the
potential is unbounded below. This raises the question as to whether these
theories have stable vacua.
It will be shown here that such $\lll>0$ supergravities in $D$ dimensions
do have  BPS domain wall solutions, i.e. solutions with
$D-1$ dimensional Poincar\' e invariance and 
which preserve half the supersymmetries, so that  these solutions are
stable vacua for such theories.

More generally, there is a large class of gauged supergravity theories,
many of which do not have AdS or Minkowski space solutions.
However, it will be seen that they do  have supersymmetric domain wall
solutions, which are candidate groundstates. They typically also admit
supersymmetric electrovac or magnetovac solutions, which are product
space solutions with an electric or magnetic flux on one of the two
factors [\evacs].

There is then a  class of supersymmetric models 
in $D=5$ which have both a $D=5$ de Sitter solution and a BPS domain wall
solution; these include the $SO(3,3)$ gauged $N=8$ theory, but there
are also models with less supersymmetry and more adjustable parameters.
If there are such models in which the BPS solution 
admits a brane-world interpretation giving   effective 4-dimensional
physics as in [\randsum], there is the interesting possibility
of cosmological models which could have a phase  of 5-dimensional
inflation followed by a transition to a brane-world scenario with
effective four-dimensional physics. The models that will be
discussed here do not  fit in with    such an
interpretation, but the possibility of finding models with bulk inflation
followed by a transition to a brane-world phase seems worth pursuing.

The model that will be analysed here is  $D$-dimensional gravity coupled to a
single scalar
$\phi$, with Lagrangian
$$
{\cal L} = \sqrt{-\det g}\left[{1\over2} R - {1\over2}(\partial\phi)^2
-V(\phi)\right]
\eqn\lag$$
Of particular interest will be  
truncations of supergravities down to such   models, with 
the truncations   chosen such that
critical points of $V(\phi)$ are also critical points of the full
supergravity potential,  by an argument of  Warner [\Warn].
Following [\Sken], 
the potential will be taken to be of the form
$$
V= 2(D-2)\left[(D-2)(w')^2 - (D-1)w^2\right]
\eqn\vis$$
for some \lq superpotential' $w(\phi)$.
The domain-wall ansatz for the metric is
$$ds^2= e^{2 A(r)} ds^2\left(\bE^{(1,D-2)}\right) + dr^2
\eqn\met$$
with scalar field $\phi(r)$ depending only on the transverse coordinate
$r$. 
Such solutions can be interpreted as representing renormalization group
flows, with monotonic {\cal C} -function [\cfun,\Freed,\Sken]
$${\cal C} = {\cal C}  _0/\left[\partial_r A(r)\right]^{D-2}
\eqn\abc$$
Critical points of the potential $V$  correspond to RG fixed points,
but there are domain wall solutions even for potentials $V$ without
critical points.

The domain wall solutions of   [\Sken] are 
such configurations 
satisfying the
following pair of first-order equations:
$$\eqalign{
\partial_r A &= \mp 2 w(\phi) \cr
\partial_r \phi  &= \pm 2(D-2) w'(\phi)
\cr}
\eqn\steq$$
with one or other choice of sign.
The second-order equations following from \lag\ are then satisfied.
These equations were found by seeking 
  solutions which extremise the   energy
$$E[A,\phi] = {1\over 2}\int_{-\infty}^\infty\! dr\, 
e^{(D-1)A}\left[(\partial_r\phi)^2 -
(D-1)(D-2)(\partial_r A)^2 + 2V\right]\, .
\eqn\abc$$

The equations \steq\ also follow from demanding the existence of spinors
satisfying
$$(D_m + w\Gamma_m)\epsilon =0\eqn\ksa$$
and 
$$
\left[\Gamma^m\partial_m\phi - 2(D-2)w'\right]\epsilon =0
\eqn\ksb$$
For solutions satisfying \steq, the
spinors satisfying \ksa,\ksb\ are
$$\ee = e^{A/2} \ee_0\eqn\spin$$
 with $\ee_0$ a constant spinor satisfying 
$$\ggg_r \ee_0 = \pm \ee_0\eqn\chir$$

For a supergravity theory in a background with the only non-vanishing
fields  being the metric and a single scalar $\phi$, there will be a
supersymmetry of the background for each  spinor $\ee$ satisfying
the Killing spinor conditions.
The condition from the vanishing of the gravitino variation
is typically of the form
$$D_m \epsilon ^a+ \Gamma_mW^a{}_b\epsilon ^b =0
\eqn\ksc$$
where $a=1,2,...,N$ labels the supersymmetries and 
$W^a{}_b(\phi) $ is a scalar-dependent matrix.
If one of its eigenvalues is $w(\phi)$ with multiplicity $m$, then there
are  Killing spinors corresponding 
to the solutions \spin\ of \ksa, provided that $\phi$ can be chosen to
satisfy the equation arising from the vanishing of the variation of
the spin-1/2 fields. If so, then the background will preserve at least a 
fraction 
$m/2N$ of the supersymmetry.

The extra condition on a spinor satisfying \ksa\ from the vanishing of the
variation of the spin-1/2 fields
is often of the form
$$
\left[\Gamma^m\partial_m\phi -Y\right]\epsilon =0
\eqn\sdfdsj$$
for some $Y(\phi)$.
Then the integrability conditions from \ksa,\sdfdsj\ are consistent with
the field equations from a lagrangian of the form \lag\ only if
$$Y=2(D-2)w'
\eqn\abc$$
and the potential takes the form \vis.
The examples that will be explored in later sections 
have a diagonal matrix
$$W_{ab}=w\dd_{ab}\eqn\Wis$$
and the backgrounds satisfying \steq\   preserve half the
supersymmetry.

In section 2, truncations of certain supergravity models will be studied
and found to have Killing spinor conditions which are precisely of
 the form \ksa,\ksb, so that there are BPS domain walls corresponding to
solutions of \steq.
In particular, all of the non-compact gaugings of $N=8$ supergravity,
including those with potential of the form \lampot, and the $N=4$ gauged
supergravity of [\gat]
 will be shown to have BPS   domain wall
solutions.
The superpotential in all these supergravity cases is of the form  
$$w=
c_1 e^{-  a_1\phi } + c_2 e^{  a_2 \phi }
\eqn\wans$$
for some constants $a_1,a_2,c_1,c_2$.

Domain wall solutions of supergravities, and in particular of those with 
superpotentials of the form \wans, have been extensively studied
[\Sken,\Freed-\Behrn]. If $x_1=0$ or $x_2=0$,
then the potential is an exponential, and the domain wall solutions were
found in [\stains,\LPS,\CowdallTW].
The case in which both $x_1,x_2$ are non-zero and $a_1\ne a_2$
has been studied in [\LPS,\Sken].
The case in which both $x_1,x_2$ are non-zero and $a_1=  a_2$
has received little attention, however, and this is the case 
that arises in some of the gauged supergravities 
with de Sitter solutions.
In section 3, the domain wall  solutions for potentials of the form \wans\
will be discussed.
While this paper was in preparation, the paper   [\Ahn] 
appeared which has
some overlap with the results presented here.

\chapter{Gauged Supergravity} 

\section {$N=4$, $D=4$ Gauged Supergravity}

The ungauged $N=4$ supergravity in $D=4$ has a global $SU(4)\times
SL(2,\R)$ symmetry and a local $U(4)$ symmetry.
The  $SU(4)\times
SL(2,\R)$
  is a duality symmetry of the equations of motion and only
a   $SU(2)\times SU(2)\times
SO(1,1)$    subgroup is a symmetry of the action.
The bosonic sector consists of 6 vector fields 
transforming as a ({\bf 6,1}) of $SO(4)\times
SO(1,1)$  
and
a complex scalar $\phi$, taking values in  the coset
$SL(2,\R)/U(1)$.
Gauging consists of promoting
the rigid $SU(2)\times SU(2)$ symmetry to a local one with
coupling constants $g_1,g_2$ for the two $SU(2)$ factors,
with the 6 vector fields becoming the gauge fields, and adding
$g_1,g_2$-dependent terms, including a scalar potential, to obtain a
supersymmetric theory.
The scalar potential of [\gat] can be written in the form [\ncto]
$$
V= - \2 \left[
(g_1^2+g_2^2)\cosh (2\vert \phi \vert) + 4g_1g_2
+(g_1^2-g_2^2){Re(\phi)\over  \vert \phi \vert}
\sinh (2\vert \phi \vert)
\right]
\eqn\abc$$
Field redefinitions bring the theory to one 
of three distinct cases with $\xi=g_2/g_1$ being $1,-1$ or $0$
[\Zcoup,\ncto]. First, if $g_1=g_2$, one obtains the gauging of [\nfor]
with a  potential
$$V=-\2 g^2(\cosh (2\vert  \phi \vert  ) +2)
\eqn\potl$$
and there is  a supersymmetric AdS solution with
$$\lll=-{3\over 2}g^2
\eqn\abc$$
If $g_1=-g_2$, one obtains the gauging of [\gat]
with a $\theta$-independent potential
$$V=-\2 g^2(\cosh (2\vert  \phi \vert  ) -2)
\eqn\pota$$ with a maximum at $\phi=0$  so that there is a
non-supersymmetric de Sitter solution with positive cosmological constant
$$\lll=\2 g^2
\eqn\abc$$
Finally, if $g_2=0$, one obtains the $SU(2)$
gauging of [\FreedS].
These three theories can be obtained as consistent truncations of
the $N=8$ theories with gauge groups $SO(8)$, $SO(4,4)$ and $CSO(4,4)$
respectively, to be reviewed in the next section.

Setting $Im(\phi)=0$, the dependence on $\vvv=\sqrt {2}Re(\phi)$ can
be written as
$$V = 
  - g^2 \left( e^{\sqrt {2}\vvv}+4\xi
 + \xi^{2} e^{-\sqrt {2}\vvv}\right)
\eqn\abc$$
and this can be written as \vis\ with superpotential
$$
w={1\over 2\sqrt {2}}g(e^{{1\over \sqrt {2}}\vvv}+\xi e^{-{1\over \sqrt
{2}}\vvv}) \eqn\abc$$
 and the Killing spinor conditions 
reduce to \ksa,\ksb\ if the only non-zero matter field is $\vvv$.

\section {$N=8$, $D=4$ Gauged Supergravity}

The ungauged $N=8$ supergravity in $D=4$ has a global $E_{7(7)}$ symmetry
and a local $SU(8)$ symmetry [\CJ].
The
$E_{7(7)}$ is a duality symmetry of the equations of motion, but there is
an $SL(8,\R)$ subgroup which is a symmetry of the action.
The bosonic sector consists of 28 vector fields 
transforming as a {\bf 28} of $SL(8,\R)$
and
70 scalars, taking values in  the coset
$E_7/SU(8)$.
Gauging consists of promoting a 28-dimensional subgroup $K$
of $SL(8,\R)$ to a local symmetry. The 28 vector fields
become the gauge bosons, so that it is necessary that the subgroup $K$ is
chosen so that the  {\bf 28} of $SL(8,\R)$ becomes the adjoint of $K$.
Then supersymmetry requires the addition of terms depending on the
coupling constant
$g$   to the action and supersymmetry transformation rules, including a
scalar potential proportional to $g^2$.
In [\dwn], the gauging with $K=SO(8)$ was constructed, and in [\nct-\nctt]
gaugings were   constructed with non-compact gauge groups
$K=SO(p,8-p)$ or
the non-semi-simple gauge groups
$CSO(p,q,r)$   
for all non-negative integers $p,q,r$ with
$p+q+r=8$.
Here
 $CSO(p,q,r)$ is 
the group contraction of $SO(p+r,q)$
 preserving a symmetric metric with $p$ positive eigenvalues, $q$
negative ones and $r$ zero eigenvalues. Then  $CSO(p,q,0)=SO(p,q)$ and
$CSO(p,q,1)=ISO(p,q)$. 
The Lie algebra of $CSO(p,q,r)$ is
$$
[ L_{ab}, L_{cd} ] = L_{ad} \eta_{bc} -L_{ac} \eta_{bd} -
L_{bd} \eta_{ac} +L_{bc} \eta_{ad}
\eqn\alg$$
where
$$ \eta_{ab} =\pmatrix{
{\bf 1}_{p \times p} & 0 &0 \cr
0 & {\bf 1}_{q \times q} &0\cr 0& 0& 0 _{r\times r}
}\eqn\abc$$     
  $a, b = 1, \cdots, 8$ and $L_{ab}=-L_{ba}$. 
Note that despite the non-compact
gauge groups, these are unitary theories, as the vector kinetic term is
not the minimal term
 constructed with the indefinite Cartan-Killing metric,  but is
constructed  
with a positive definite scalar-dependent matrix. The $CSO(p,q,r)$
gauging and the $CSO(q,p,r)$
gauging are equivalent.
 In [\frefo], it was argued that these are the only
possible gauge groups.

The 70 scalars consist of 35 scalars 
  parameterising the coset $SL(8,\R)/SO(8)$ and which can be
represented by an $8\times 8$ unimodular matrix $S$, and 35 pseudo-scalars
  parameterised by
an anti-self-dual 4-form of  $SL(8,\R)$.
Let $K_{p,q,\xi}$ be the subgroup of
$SL(8, {\R})$ whose algebra is \alg\ with
$$ \eta_{ab} =\pmatrix{
{\bf 1}_{p \times p} & 0   \cr
0 & \xi {\bf 1}_{q \times q} 
}\eqn\abc$$  
parameterised by $\xi$, with $p+q=8$.
For $\xi =1$, this is $SO(8)$, for $\xi=-1$ this is $SO(p,q)$ and
for $\xi=0$ this is the non-semi-simple 
$CSO(p,0,q)$, which it is convenient to abbreviate to
$CSO(p,q)$.
For other $\xi >0$ ($\xi<0$), this is isomorphic to the algebra with
$\xi =1$ ($\xi =-1$).

In [\nctt], the potential $V(s)$
  of the $N=8$ theory with
gauge group  $K_{p,q,\xi}$ ($p+q=8$) was found explicitly for the
case in which there is only one non-zero scalar $s$, which is   the
$SO(p)\times SO(8-p)$ invariant
scalar represented by the $SL(8,\R)$ matrix
$$ S =\pmatrix{
e^{-s}{\bf 1}_{p \times p} & 0   \cr
0 &e^{ps/q} {\bf 1}_{q \times q} 
}\eqn\abc$$ 
The potential is [\nctt]
$$V_{p, q, \xi}  =   
-{1 \over 32} g^2 e^{2s}\left(  2p(q-3p) - 16pq \xi
e^{-8s/q} + 2q(p-3q)\xi^{2}e^{-16s/q}\right)
\eqn\potentiala$$
Explicitly, this gives [\nctt]
$$\eqalign{
V_{7, 1, \xi} & =   
{1 \over 8} g^2 \left( -35e^{2s}- 14\xi
e^{-6s} + \xi^{2}e^{-14s}\right), 
\cr 
V_{6, 2, \xi} & =  -3 g^2 \left( e^{2s}+ \xi e^{-2s}\right), 
\cr 
V_{5, 3, \xi} & = 
  -
{3 \over 8} g^2 \left( 5e^{2s}+ 10\xi
e^{-2s/3} + \xi^{2}e^{-10s/3}\right), 
\cr 
V_{4, 4, \xi} & = 
  - g^2 \left( e^{2s}+4\xi
 + \xi^{2} e^{-2s}\right), 
\cr 
V_{3, 5, \xi} & = 
  -
{3 \over 8} g^2 \left( e^{2s}+ 10\xi
e^{2s/5} + 5\xi^{2} e^{-6s/5}\right), 
\cr 
V_{2, 6, \xi} & = 
  -3 g^2 \xi \left( e^{2s/3}+\xi
e^{-2s/3} \right), 
\cr 
V_{1, 7, \xi} & = 
  {1 \over 8} g^2 \left( e^{2s}- 14\xi
e^{6s/7} -35 \xi^{2}e^{-2s/7}\right). 
\cr }
\eqn\potential$$
From the argument of [\Warn], any critical point of these potentials
$V(s)$ is also a critical point of the full potential. 
The $SO(8)$ gauging has a critical point of $V_{p, q, 1}$ at $s=0$
preserving $SO(8)$ and all 32 supersymmetries. The other critical points
of these potentials  break  all supersymmetries and are as follows
[\nctt]. The potential $V_{7, 1, 1}$ has two critical points and so the
$SO(8)$ gauging has two $SO(7)$-invariant critical points with $\lll <0$, 
the one at $s=0$, and
one which breaks all supersymmetries.
The potential for $SO(4,4)$ is of the form \lampot\ and has a $\lll >0$
critical point at $s=0$.
The $SO(5,3)$  gauging (which is  equivalent to the $SO(3,5)$ gauging)
also has a
  $\lll >0$
critical point, at $s=-{3\over 8} \log 3$, and the potential is of the
form \wans\ with $a_1\ne a_2$.
When $\xi=0$, these potentials are all of the form
$$V \propto  - e^{2s}
\eqn\abc$$
 and in the case $p=2$, the potential 
$V_{2, 6, 0}$ vanishes identically, and there is a $\lll=0$
critical point of the full potential, so that there is a
non-supersymmetric Minkowski space solution.

The scalar field $s$ can be written in terms of a canonically normalised
scalar $\phi$
by 
$$\phi= \sqrt{{2p\over q}} s
\eqn\abc$$
In the Killing spinor condition \ksc, the tensor $W^{ab}$ is given by the
so-called $A_1^{ab}$ tensor of [\nct-\nctt], and 
for a background 
of the $K_{p,q,\xi}$ gauging 
with $\phi$ the only non-vanishing matter field, this is of the form
\Wis\ with [\nctt]
$$
w_{p,q,\xi}(  \phi) =   {\sqrt{2} g \over 8} \left( p e^{\sqrt{ {q \over 2p}}
\phi} + q \xi  e^{-\sqrt{ {p \over 2q}}\phi} 
\right) 
\eqn\wpq$$
Then the Killing spinor conditions
are
precisely \ksa,\ksb\ with $w$ given by \wpq\
and it is  straightforward to check that the 
potential \potentiala\ is given in terms of this superpotential
$
w_{p,q,\xi}$ by \vis\ with $D=4$.

\section {$N=8$ Gauged Supergravity in  $D=5$ and $D=7$.}

The ungauged $N=8$ supergravity in $D=5$ has an action invariant under a
global
$E_{6(6)}$ symmetry and a local $USp(8)$ symmetry [\Cremmer].
The bosonic sector consists of 27 vector fields 
transforming as a {\bf 27} of $E_6$
and
42 scalars, taking values in  the coset
$E_6/USp(8)$.
It has a dual form with only
$SL(6,\R)\times SL(2,\R)$ global symmetry
in which 12 of the vector fields are dualised to 2-forms, leaving
15 vector fields transforming as the ({\bf 15,1}) of $SL(6,\R)\times
SL(2,\R)$.
 The 
gaugings arise from making a 15-parameter subgroup $K$ of 
$SL(6,\R)$ local, with the 15 vectors becoming the gauge bosons.
The gaugings with
$K=SO(6)$ [\PPV,\gunwar]  and $K=SO(p,6-p)$     [\gunwar]
arise in this way, but non-semi-simple gaugings arise from a slightly
different construction [\fref] and will not be discussed here.

Twenty of the 42   scalars 
  parameterise the coset $SL(6,\R)/SO(6)$ and  can be
represented by a $6\times 6$ unimodular matrix $S$.
For the $SO(6)$   or $SO(p,q)$   gauging with $p+q=6$,
consider the
$SO(p)\times SO(6-p)$ invariant
scalar $\phi(x)$ represented by the $SL(6,\R)$ matrix
$$ S =\pmatrix{
e^{-a\sqrt{{p\over q}}\phi }{\bf 1}_{p \times p} & 0   \cr
0 &e^{ a\sqrt{{q\over p}} \phi} {\bf 1}_{q \times q} 
}\eqn\sis$$
as in [\gunwar], where $a$ is a normalisation.
Then for a scalar background with $\phi(x)$ the only-non-vanishing matter
field,
  the Killing spinor conditions
for a background 
of the  $SO(6)$ ($\xi=1$) or $SO(p,q)$ ($\xi=-1$)
 gauging 
  are
precisely \ksc, \Wis\  with $w$ proportional to  $tr( \eta S)$  where
$\eta$ is the $SO(p,q)$ invariant metric [\gunwar]
so that
$$
w_{p,q,\xi}(  \phi) \propto g \left( p e^{a\sqrt{ {q \over p}}
\phi} + q \xi  e^{-a\sqrt{ {p \over q}}\phi} 
\right) 
\eqn\wisss$$

The ungauged $N=8$ supergravity in $D=7$ has  
a rigid $SL(5,\R)$ symmetry and a local $SO(5)$ symmetry, with
scalars taking values in
$SL(5,\R)/SO(5)$. In [\PPVa], it was shown that one can gauge an $SO(p,q)$
subgroup of the $SL(5,\R)$ symmetry  for any $p+q=5$. For the $SO(5)$   or
$SO(p,q)$   gauging with $p+q=5$,   the
$SO(p)\times SO(5-p)$ invariant
scalar $\phi(x)$ can be represented by the $SL(5,\R)$ matrix \sis.
 The Killing spinor conditions
for such a background 
of the  $SO(5)$ ($\xi=1$) or $SO(p,q)$ ($\xi=-1$)
 gauging 
  are again
  \ksc, \Wis\  with $w$ again of the form \wisss.

\chapter{Solutions}

\section{The ansatz}

The metric \met\ can be brought to the  form
$$ds^2= e^{2 A(r)} ds^2\left(\bE^{(1,D-2)}\right) +e^{2B} dr^2
\eqn\abc$$
for any function $B(r)$ by a coordinate transformation
$r \to f(r)$ with $f'= e^{-B}$.
The equations \steq\ then become
$$\eqalign{
\partial_r A &= \mp 2 w(\phi) e^B\cr
\partial_r \phi  &= \pm 2(D-2) w'(\phi)e^B
\cr}
\eqn\steqb$$
A useful choice is   
$$B=\ll A\eqn\abc$$
 for some constant $\ll$, as this simplifies the equations in some cases.
Choosing basis 1-forms 
$$e^\mu=e^Adx^\mu,\qq  e^r=e^{\ll A}dr
\eqn\abc$$
the curvature 2-form $\Theta^{MN}$ has the frame components
$$
\eqalign{
\Theta^{r \mu}&=(-A''+(\ll -1)(A')^2)e^{-2\ll A}e^r\wedge e^\mu\ , \cr
\Theta^{\mu\nu}&=-(A')^2e^{-2\ll A}e^\mu\wedge e^\nu\ 
\cr}
\eqn\abc$$

Consider the case in which the superpotential
is of the form
$$w=
c_1 e^{-  a_1\phi } + c_2 e^{  a_2 \phi }
\eqn\abc$$
for some constants $a_1,a_2,c_1,c_2$, with $a_1,a_2> 0$. All of the
supergravity theories discussed in the last section have 
superpotentials of this form.
There are a number of different cases, depending on whether or not
$a_1=a_2$, and these will be considered separately.

\section{ Generic Case: $a_1\ne a_2$}

A solution of the equations \steqb\ was found in 
[\LPS].
It is convenient to choose $B=\ll A $ with
$$\ll = {a_1 a_2 \over  (D-2) (a_2-a_1)}
\eqn\abc$$
Then \steqb\ are solved by
$$e^{ (a_1 - a_2)\phi} = r
\eqn\abc$$
and
$$e^{-\ll A} = x_1\, r^{ {a_2}\over {a_2-a_1}} + x_2\,
r^{-{{a_1}\over {a_2-a_1}}}
\eqn\abc$$
where
$$x_1 = \mp 2(D-2) a_1c_1, \qq
x_2 = \pm 2(D-2) a_2c_2
\eqn\abc$$

If $c_1=0$ or $c_2=0$, the  potential is a simple exponential 
and the domain wall solutions of [\stains,\LPS,\CowdallTW] are recovered.
In this case, either the region as $r\to \infty $ or the region as
$r\to 0 $ is asymptotically flat, but
the scalar $\phi$ is
proportional to $\log r$. 
 The regions $r\to 0 $ and $r\to \infty $ can be interchanged by the 
coordinate transformation
$r\rightarrow 1/r$, so that the asymptotically flat region can always be
arranged to be at large $r$. With the asymptotically flat region arranged to
be at large $r$, writing $r=\vert z \vert $ gives a solution 
defined for all real $z$ that is asymptotically flat as $z \to \pm \infty$ and
is symmetric under the reflection $z \to -z$ and has a singular domain wall at
$z=0$.

If both $c_1,c_2$ are non-zero and have opposite sign, the frame
components of the curvature  diverge both as $r\to 0 $ and as $r\to
\infty $ [\LPS].
Such models arise from the $SO(p,q)$ gauged supergravities in $D=4,5$ with
$p\ne q$ and $\xi=-1$

If $c_1,c_2$ are non-zero and have the same sign, then
$e^{-\ll A}$   vanishes at $r=r_c$ where
$$ r_c^{ {a_2+a_1}\over {a_2-a_1}}= {a_2c_2 \over a_1 c_1}
\eqn\abc$$
and it is necessary to restrict to the region in which $e^{-\ll A}$ is
positive. 
Suppose the signs are such that the solution is restricted to the region
$0\le r\le r_c$. 
The solution is singular at $r=0$, but the curvature vanishes at $r=r_c$ and
the scalar field is finite there, so    there is a flat region near
$r=r_c$,  which is at an infinite distance from any point with $r<r_c$.
Again, writing $r=\vert z \vert$ gives a solution
defined for real $\vert z \vert < r_c$ which is asymptotically flat as $ z \to
\pm r_c$ and which has a singular domain wall at $z=0$. Such models arise from
the  $N=8$ theories with compact gauge groups, with $\phi$ the $SO(p)\times
SO(q)$ invariant scalar, with $p\ne q$.

\section{ Special Case: $a_1= a_2$ and $c_1= -c_2$}

In this case
$$w= c \sinh (a\phi)
\eqn\abc$$
Taking $\ll=0$ and setting
$\varphi = a \phi$,
\steq\ gives
$$
\pa _r \vvv
= \aa \cosh \vvv
\eqn\abc$$
where
$$ \aa= \pm 2(D-2) c a
\eqn\abc$$
This can be integrated to give
$$
\aa (r-r_0) = gd (\vvv)
\eqn\ris$$
where $gd (\vvv)$ is the {\it Gudermannian} defined by
$$gd(x) =\int _0 ^x{dt \over \cosh t}
\eqn\abc$$
and can be written as
$$ gd (x) = 2 arctan (e^x) -{\pi \over 2}
\eqn\abc$$
It enjoys many remarkable properties, such as
$$ \eqalign{
\tanh (x) &= \sin (y) \cr
 \sinh (x) &= \tan (y) \cr
\cosh (x) &= \sec (y) \cr}
\eqn\gud$$
where $y=gd(x)$.
Then \ris,\gud\ imply 
$$\sinh (\vvv) = \tan [\aa (r-r_0)]
\eqn\abc$$
or, after a shift of $r$ to absorb constants of integration,
$$ e^\vvv = \tan (\aa r'/2)\eqn\abc$$
with $r'=r-r_0 +\pi/(2\aa)$.
Then $A$ satisfies
$$\pa_r A= \mp 2c \sinh \vvv =\mp 2c\tan [\aa (r-r_0)]
\eqn\abc$$
and so 
$$ e^A = \left( \cos [\aa (r-r_0)] \right) ^{\pm 2c/\aa}
\eqn\abc$$

Note that in this solution $r$ is restricted to the region
$$
\vert \aa (r-r_0) \vert < {\pi\over 2}
\eqn\abc$$
as  at $\vert \aa (r-r_0) \vert = {\pi\over 2}$
the scalar field $\vvv$ becomes infinite
and $e^A$ either vanishes or diverges, depending on the choice of sign in
\steq.
However, for either sign
$$
\Theta^{\mu\nu}=-4c^2 \tan ^2 (\aa (r-r_0))
e^\mu\wedge e^\nu\ 
\eqn\abc$$
so that the frame components of the curvature 
diverge at the boundaries $\vert \aa (r-r_0) \vert = {\pi\over 2}$.

\section{ Special Case: $a_1= a_2$ and $c_1= c_2$}

In this case
$$w= c \cosh (a\phi)
\eqn\abc$$
Taking $\ll=0$ and setting
$\varphi = a \phi$,
\steq\ gives
$$
\pa _r \vvv
= \aa \sinh \vvv
\eqn\abc$$
where
$$ \aa= \pm 2(D-2) c a
\eqn\abc$$
This can be integrated to give
$$
e^{\aa (r-r_0)} =   \tanh (\vvv/2) =\sqrt{{ \cosh \vvv -1 \over
\cosh \vvv +1}}
\eqn\abc$$
so that 
$$\cosh \vvv = f [\aa (r-r_0)]
\ek
where
$$f(y)= {1+e^{2y} \over 1-e^{2y}}\eqn\abc$$
Then $A$ satisfies
$$\pa_r A= \mp 2c \cosh \vvv =\mp 2c
f [\aa (r-r_0)]
\eqn\abc$$ 
which implies
$$A =\mp 2c \left( (r-r_0)- \aa^{-1} \log \left[ 1-e^{2\aa (r-r_0)}
\right]\right)
\eqn\abc$$
There is a singularity at $r=r_0$ where $\vvv,A$ and the curvature 2-form
diverge.
If $\aa>0$, the solution is restricted to the region $r<r_0$ and
if $\aa<0$, it is restricted to the region $r>r_0$.
The solution is non-singular as $r\to   \infty $ ($\aa<0$) or
$r\to -  \infty $ ($\aa>0$).

\section{ Special Case: $a_1= a_2=a$ and $c_1\ne c_2$}

Here
$$w=
c_1 e^{-  a\phi } + c_2 e^{  a \phi }
\eqn\wans$$
and there are two cases, depending on the relative signs of $c_1,c_2$. If
$c_1c_2<0$, then the solution is
$$
\eqalign{
e^{a\phi} &=
\sqrt{{-c_1\over c_2}} \tan \rho
\cr
e^{-A}  &=\2 \sin (2\rho)
\cr}
\ek
where
$$\rho= 2(D-2)a\sqrt{-c_1c_2}(r-r_0)
\ek

If
$c_1c_2>0$, then the solution is
$$
\eqalign{
e^{a\phi} &=
\sqrt{{c_1\over c_2}} {e^ \rho -1 \over e^ \rho +1}
\cr
A &=\pm
{2\over (D-2)a\sqrt{c_1c_2}}\left[ \rho -
\log\left( e^{2\rho}-1\right)\right]
\cr}
\ek
where
$$\rho= 2(D-2)a\sqrt{c_1c_2}(r-r_0)
\ek

\refout
\bye